\documentclass[sigconf,authorversion,nonacm]{acmart}
\AtBeginDocument{%
  }

\usepackage{glossaries}
\makeglossaries

\begin{document}

\title{An Exemplar of a Digital Twin in Mechanical Engineering: Understanding Model Hybridization}

%

\author{Mahussi Datongnon}
\affiliation{%
  \institution{Centre Inria d’Université Côte d’Azur}
  \city{Sophia-Antipolis}
  \country{France}}
\email{mahussi-jeff-fidele.datongnon@inria.fr}
\orcid{1234-5678-9012}

\author{Hubert LeJeune}
\affiliation{%
  \institution{CETIM}
  \city{Nantes}
  \country{France}}
\email{Hubert.Lejeune@cetim.fr}

\author{Yoann Jus}
\affiliation{%
  \institution{CETIM}
  \city{Nantes}
  \country{France}}
\email{Yoann.Jus@cetim.fr}

\author{Benoit Combemale}
\affiliation{%
 \institution{Université de Rennes/Inria/Irisa, team DiverSE}
 \city{Rennes}
  \country{France}}
\email{benoit.combemale@inria.fr}

\author{Julien Deantoni}
\affiliation{%
 \institution{Université Côte d’Azur/I3S/Inria, team Kairos}
 \city{Sophia-Antipolis}
  \country{France}}
\email{julien.deantoni@univ-cotedazur.fr}




\newacronym[plural=DTs]{dt}{DT}{Digital Twin} 

\newacronym{dts}{DTs}{Digital Twins}

\newcommand{\cetim}[0]{Cen\-tre Tech\-ni\-que des In\-dus\-tries Mé\-ca\-ni\-que (CETIM)}

\newcommand{\fmu}[0]{Functional Mock-up Unit (FMU)}

\newcommand{\fmi}[0]{Functional Mock-up Interface (FMI) }

\newcommand{\ssp}[0]{System Structure and Parameterization (SSP) }

\begin{abstract} \label{abstract}
   

\Glspl{dt} are widely adopted across a variety of application domains \cite{jonesCharacterisingDigitalTwin2020}. In industrial sectors, particularly in mechanical engineering, they accelerate product development, reduce risks, enable early issue prediction, and lower sustainment costs \cite{ferrariDigitalTwinsMechanical2024}. In practice, \glspl{dt} increasingly integrate physics-based (deductive) and data-driven (inductive) models into hybrid models \cite{wangHybridPhysicsbasedDatadriven2022a} combining the complementary strengths of both modeling paradigms. In this paper, we refer to this integration paradigm as \emph{hybridization}. Despite this trend, the engineering of hybrid DTs \cite{garroDesigningISO23247compliant2026a} that is, remains insufficiently documented. Hybridization is often introduced in an ad hoc manner, and its implementation is only partially made explicit, which limits reproducibility and transferability.
This paper reports on the development of an existing fluidic loop digital twin at \textit{\cetim}. The DT is described using the characterization framework of \cite{gilSystematicReportingFramework2025}, providing a structured view across its lifecycle dimensions. To make hybridization explicit, the case is further analyzed through a complementary characterization structured along two dimensions: motivation and realization. The resulting description provides a traceable account of hybridization decisions and supports the documentation and transfer of hybrid DT engineering practices.

\end{abstract}






\maketitle

\section{Introduction}\label{introduction}
A \gls{dt} can be defined as a complex software system that intentionally represents an original system, capable of reflecting changes in that system, interacting with it, and providing operational value \cite{vicat-blancFabriqueJumeauxNumeriques2026}. \gls{dts} have emerged as key components of industrial systems, enabling monitoring, simulation, analysis, and decision support. \gls{dt} increasingly combine physics-based (deductive) models with data-driven (inductive) approaches, leading to hybrid systems that leverage complementary modeling capabilities.

Despite this evolution, the engineering of hybrid DTs remains insufficiently formalized. In practice, hybridization is often introduced in an ad hoc manner, and both its motivation and implementation are only partially documented. As a result, essential engineering rational and decisions such as model integration, validation, and interaction remain implicit, leading to limited transferability across projects and application domains.

This paper addresses this gap through an industrial case study conducted at CETIM, focusing on the development of an existing fluidic loop \gls{dt}. It presents a reusable engineering exemplar based on the development of the fluidic loop \gls{dt}, a structured characterization of the \gls{dt} using a systematic reporting framework \textit{{\cite{gilSystematicReportingFramework2025}}}~\cite{gilSystematicReportingFramework2025}, and a complementary hybridization perspective organized along two dimensions (motivation and realization). In addition, it highlights transferable lessons learned, supported by supplementary material that can be retrieved in~\cite{datongnon_parameter_identifier_cetim,datongnon_parameter_estimator_cetim,datongnon_parameter_validation}.

The remainder of the paper is organized as follows. Section~\ref{fluidic-loop-case-study} presents the existing fluidic loop case study. Section~\ref{dt-characterization} details the digital twin characterization following \cite{gilSystematicReportingFramework2025}. Section~\ref{hybridization-perspective} introduces the hybridization perspective followed by a discussion in Section~\ref{results-and-discussion}.

\section{Fluidic loop case study}\label{fluidic-loop-case-study}

\noindent
\begin{figure}[htbp]
    \vspace{0pt}
    \centering
    \includegraphics[width=\linewidth]{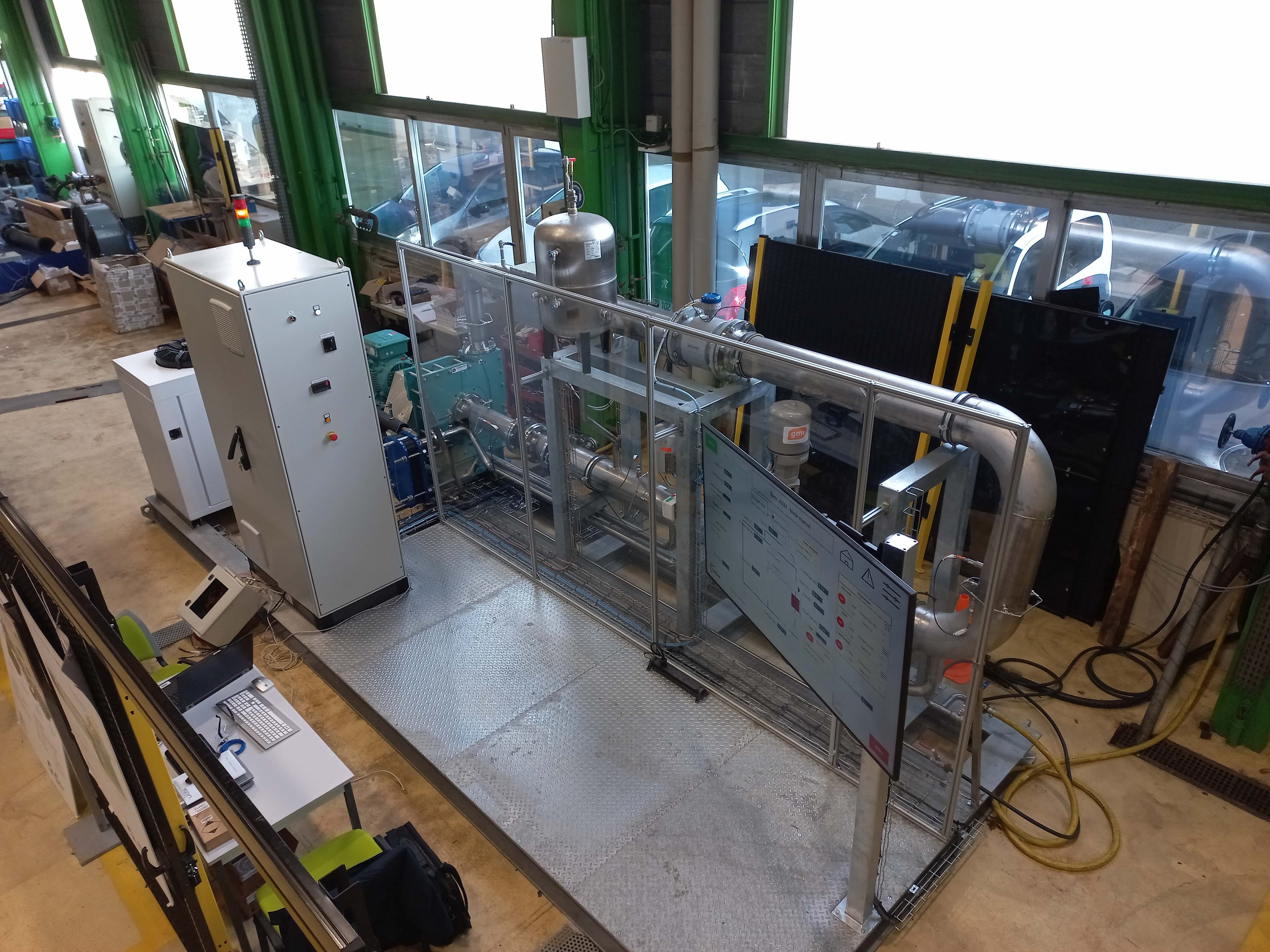}
    \captionof{figure}{Fluidic loop.}
    \Description{Fluidic loop (physical entity).}
    \label{fig:fluidic-loop}
\end{figure}

This section presents an industrial case study based on a fluidic loop and its associated \gls{dt}. The physical system is illustrated in Figure~\ref{fig:fluidic-loop}. The \gls{dt} is designed to provide monitoring and decision-support capabilities. The physical system consists of a closed hydraulic loop equipped with a centrifugal pump, a heat exchanger, and several control elements and actuators. Its function is to deliver the flow rate, pressure, and/or temperature requested by the operator.

\subsection{DT characterization}\label{dt-characterization}
The \gls{dt} of the fluidic loop is characterized using a systematic reporting framework \cite{gilSystematicReportingFramework2025}. A complete description of the \gls{dt} according to this framework is available in ~\cite{edtlab_uc05_fluid_loop}. The content presented here provides a concise synthesis relevant to the present study. The framework  proposed by \cite{gilSystematicReportingFramework2025} consolidates and harmonizes existing DT characterization approaches proposed by 
\cite{daliborCrossDomainSystematicMapping2022, oakesImprovingDigitalTwin2021, gilSystematicReportingFramework2025}, resulting in a unified set of \textbf{Merged Characteristics (MCi)} \cite{gilSystematicReportingFramework2025}. These characteristics provide a systematic basis for describing both the architecture and the functionalities of the DT on four dimensions.

\subsubsection{Requirements, Conceptualization, and Design}\label{requirements-conceptualization-and-design}

\paragraph{System under study and physical components}\label{system-under-study-and-physical-components} : The system under study is the fluidic loop itself (MC1), composed of physicals actuators (MC2) and sensing components (MC3). The actuators include the pump speed, the regulation valve and the cooling unit, while the sensing
layer captures variables such as inlet/outlet pressures, pump speed,
temperature and flow rate.

\paragraph{Physical - virtual interactions}\label{physical---virtual-interactions} : Sensor data are collected and transmitted to the digital
environment and stored in a database, forming the
\textit{physical-to-virtual interface} (MC4). Conversely, the
\textit{vir\-tual-to-physical interaction} (MC5) enables the \gls{dt} to
influence the physical system through actuator commands, such as
adjusting valve opening or managing cooling strategies.

\paragraph{\glsentrylong{dt} services}\label{digital-twin-services}
The \gls{dt} provides several operational services (MC6), including \textit{cavitation detection}, \textit{virtual sensing} where flow rate is estimated in case of sensor failure, \textit{energy consumption analysis} through the simulation of alternative cooling strategies, and \textit{divergence detection}, which identifies discrepancies between the physical and virtual systems to trigger recalibration of \gls{dt} models.

\subsubsection{Realization}\label{realization}$\newline$

The realization of the \gls{dt} for this fluidic loop relies on
the integration of heterogeneous models, data pipelines, and
computational enablers, orchestrated to support the different
operational services. This section details the \gls{dt}  realization according
to models and data, enabling technologies, system orchestration,
lifecycle evolution, and fidelity considerations.

\paragraph{\glsentrylong{dt} Models and Data
(MC10)}\label{digital-twin-models-and-data-mc10} : The DT is composed of both \textit{deductive (physics-based)} and \textit{inductive (data-driven)} models, forming a hybrid modeling
architecture.

\begin{itemize}
\item
  \textbf{Physics-based models:}\\
  A 1D thermo-hydraulic simulation model (Simcenter Flomaster) is used
to represent the fluidic loop. These models simulate system variables (e.g., pressure, flow rate, temperature) based on component parameters and control inputs.
\item
  \textbf{Data-driven models:}\\
  Two inductive models complement the simulation. A \textit{parameter identifier model} (decision tree), identifying which parameters of the physical model are responsible for observed discrepancies. A \textit{fault estimation model} (LS‑SVR), estimating the corrected values of these parameters.

\item
  \textbf{Data structure:}\\
  The \gls{dt} relies on a database that stores sensor measurements and simulation data. The machine learning models are trained using \textit{synthetic datasets generated from controlled simulations of the 1D model}. The datasets are obtained by sampling control vectors and model parameters and recording the corresponding simulated process variables.
\end{itemize}

\paragraph{Tooling and Enablers (MC11)}\label{tooling-and-enablers-mc11} :
The DT implementation includes a Beckhoff Programmable Logic Controller (PLC) for automation and control, along with an industrial PC for local computation. For interoperability and simulation, the system uses the \textit{FMPy library} \cite{fmpy} to execute \fmu{} for managing simulation assets. The application layer consists of a web interface enabling visualization and user interaction. These enablers ensure data acquisition, model execution, orchestration, and interaction within the DT.

\paragraph{Digital Twin Constellation (MC12)}\label{digital-twin-constellation-mc12} :
The DT constellation follows a distributed, service-oriented architecture where sensors transmit measurements to a PLC (data acquisition and control). Data are then stored in a database and processed by DT services combining FMU-based simulations, data-driven models. Data-driven models complement the physics-based model by analyzing discrepancies and dynamically adjusting parameters at runtime.

\paragraph{Twinning Process and Digital Twin Evolution (MC13)}\label{twinning-process-and-digital-twin-evolution-mc13} :
The DT follows a progressive evolution: starting from a calibrated thermo-hydraulic 1D model, it is connected to the physical system for synchronization, then extended with core services (simulation, visualization, diagnostics). Hybridization is introduced through data-driven models enabling divergence detection and parameter correction, leading to a closed-loop adaptation where model parameters are updated. 

\paragraph{Fidelity and Validity Considerations (MC14)}\label{fidelity-and-validity-considerations-mc14} : 
The DT reproduces the behavior of the fluidic system with a fidelity that depends on both the physical model and hybridization mechanisms. The thermo-hydraulic 1D model captures the global system dynamics (pressure, flow rate, temperature) with satisfactory accuracy under nominal conditions and is initially calibrated using experimental data. Fidelity is maintained over time through dynamic adaptation, combining divergence detection between simulation and measurements with data-driven models to adjust selected parameter.

\subsubsection{Deployment}\label{deployment}$\newline$

\paragraph{Digital Twin Technical Connection (MC15)}\label{digital-twin-technical-connection-mc15} : 
The communication architecture relies on industrial protocols. Sensors and actuators interact with a Beckhoff PLC industrial I/O and internal buses. OPC UA is used for standardized data exchange.

\paragraph{Digital Twin Hosting and Deployment (MC16)}\label{digital-twin-hosting-and-deployment-mc16} :
The DT is deployed on a local industrial infrastructure combining computation, control, and visualization components. At the edge level, an industrial PC hosts DT components for low-latency processing. The data layer ensures storage of historical data and datasets locally. The application layer provides a web-based interface for supervision, visualization, and interaction with the DT.

\subsubsection{Operation}\label{operation}$\newline$
The operation phase of the DT focuses on value delivery through interaction with the physical system, combining data processing, service execution, and decision support.

\paragraph{Insights and Decision-Making (MC17)}\label{insights-and-decision-making-mc17} : 
The DT provides operational insights through several functionalities: visualization of system variables (pressure, flow rate, temperature) with comparison between measured and simulated values; detection mechanisms including cavitation alerts and divergence identification; and diagnostic capabilities to identify parameters responsible for deviations using data-driven models. The DT enables predictive analysis through what-if scenario simulations and evaluates alternative operating strategies (e.g., cooling profiles). Performance optimization is achieved via energy consumption indicators and scenario comparisons to identify the most efficient configurations.

\paragraph{Horizontal Integration (MC18)}\label{horizontal-integration-mc18} : 
The DT is integrated within both industrial and information systems. At the industrial level, it interfaces with the Beckhoff PLC for control and data acquisition, using standard protocols such as OPC UA. At the IT level, data are stored and managed locally, enabling reuse by other applications (e.g., analytics, reporting) within the industrial environment.

\paragraph{Data Ownership and Privacy (MC19)}\label{data-ownership-and-privacy-mc19}
CETIM owns both the physical test bench (located in Nantes, at CETIM premises) and the associated data. The conditions for data sharing and dissemination, as well as the definition of data formats, are currently under study.

\paragraph{Standardization (MC20)}\label{standardization-mc20} : 
The \gls{dt} relies on established standards to ensure interoperability and modularity. The \fmi standard \cite{fmi_standard} is used to execution the 1D model exported as a FMU (via  \textit{FMPy library} \cite{fmpy}). OPC-UA under IEC~62541~\cite{iec62541_19_2025}, enables standardized communication between the PLC and digital components, ensuring vendor independence.

\paragraph{\textbf{Toward Explicit Characterization of Hybridization in DTs}}

The characterization framework \cite{gilSystematicReportingFramework2025} provides a generic and structured description of \gls{dts} that is independent of a specific application domain. By organizing \gls{dts} along \textit{MCi}, it offers a common basis to describe architectures, models, data flows, services, and system interactions. However, within this framework, the relationships between heterogeneous models, in particular between deductive (physics-based) and inductive (data-driven) components, remain largely implicit. As a result, their interactions, roles, and integration strategies are often engineered in an ad hoc manner, making hybridization decisions difficult to trace and limiting knowledge transfer across DT implementations. To address this limitation, the following section introduces a complementary characterization, \textit{currently under development}, specifically dedicated for guiding and structuring the description of hybridization within the fluidic loop.

\subsection{Hybridization perspective}\label{hybridization-perspective}

To complement the DT characterization, the \textbf{Hybridization Characteristics (HC)} are defined in correspondence with the framework \cite{gilSystematicReportingFramework2025}. While this framework provides a structured description of a \gls{dt} across its lifecycle dimensions ---requirements, conceptualization and design, realization, deployment, and operation--- it does not explicitly address the rationale and implementation of hybridization. Instead, the integration of deductive and inductive models is typically distributed across several DT characteristics and remains largely implicit.

The HC are introduced to make these aspects explicit. They provide a dedicated perspective for describing hybridization, capturing both the reasons for introducing it and the way it is implemented within a DT. To this end, the HC are organized along two complementary dimensions: \textit{Motivation} and \textit{Realization}. The \textit{Motivation} dimension focuses on the need for hybridization and is primarily related to the requirements, conceptualization, and design characteristics (MC1--MC6). The \textit{Realization} dimension focuses on the implementation and operationalization of hybridization and is mainly associated with the realization and operation characteristics (MC10--MC14, MC7, and MC17). A more detailed presentation of the HC is available in~\cite{datongnon_hybridization_characteristics_motivation}

\subsubsection{Motivation dimension}\label{motivation-dimension}

The motivation dimension captures the need for hybridization, independent of its implementation, by structuring the problem context, identifying the limitations of existing DT assets, and specifying the expected, new or reinforced, capabilities of the hybridized system.

\paragraph{HC1 - Overview / Context}\label{overview-context} \textit{(Link to MC1, MC6)} : The DT relies on a 1D thermo-hydraulic simulation model supporting services such as flow simulation, cavitation detection, and energy analysis. This model is the core deductive component of the DT. 

\paragraph{HC2 - Problem Statement}\label{problem-statement} \textit{(Link to MC6, MC14)} : The simulation model is calibrated once at creation, with fixed parameters during operation. As the physical system evolves (e.g., degradation, changing conditions), discrepancies emerge between simulated and measured behaviors, leading to reduced service reliability (MC6) and decreasing fidelity (MC14). This situation raises three coupled challenges: (i) \textit{parameter identification}, i.e., determining which model parameters are responsible for observed discrepancies; (ii) \textit{parameter estimation}, i.e., inferring updated values for these parameters under varying operating conditions; and (iii) \textit{parameter validation}, i.e., detecting that any proposed update preserves consistency with the physical system and does not degrade \gls{dt} fidelity. Overall, the problem is to maintain alignment between the \gls{dt} and the physical system under evolving conditions without repeated manual recalibration.

\paragraph{
\textit{The hybridization witin this \gls{dt} consists of three complementary hybridization, each addressing one of the challenges identified in HC2 (\emph{Problem Statement}): parameter identification, parameter estimation, and parameter validation. These hybridization mechanisms are described in detail in~\cite{datongnon_parameter_identifier_cetim, datongnon_parameter_estimator_cetim, datongnon_parameter_validation, datongnon_hybridization_characteristics_motivation}. For the sake of readability, the characterization presented below focuses on the first hybridization level, namely the parameter identification} problem.}

\paragraph{HC3 - Existing Twin Assets and Gaps}\label{existing-twin-assets-and-gaps}

\textit{(Link to MC10, MC14)} : The DT relies on a 1D physics-based simulation model initially calibrated against the real system at creation time, with static parameters during operation. From a \textit{parameter identifier} perspective, no mechanism exists to dynamically identify which parameters are responsible for discrepancies observed during operation.

Overall, the DT lacks integrated capabilities for consistent parameter identifier, estimation, and validation, limiting its adaptability and robustness under evolving conditions.

\paragraph{HC4 - Intended Deliverable Capability (IDC)}\label{intended-deliverable-capability-idc} \textit{(Link to MC6, MC10, MC12)} :  
For \textit{parameter identification}: the IDC is to determine which parameter is responsible for a discrepancy observed between the physical system and its \gls{dt} counterpart.

\paragraph{HC6 - Conceptual Requirements}\label{conceptual-requirements} \textit{(Link to MC10, MC13, MC14)} 
The model must identify the parameter responsible for a discrepancy within the limits of observability defined by the available measurements and operating conditions; identification shall ensure that parameter contributions are distinguishable, enabling attribution of discrepancies to a reduced set of candidate parameters.

\paragraph{HC7 - Deliverable Qualities}\label{deliverable-quality}
\textit{(Link to MC11, MC12, MC17)} Quality expectations are defined per function. For \textit{parameter identification and estimation}, outputs must ensure low latency and stable estimates, remaining compatible with DT real-time services without disrupting simulations.

\subsubsection{Realization Dimension}\label{realization-dimension}
The realization dimension describes how hybridization capabilities are implemented and executed within a \gls{dt}. It captures both the design-time and runtime realization of hybridization deliverables, including their orchestration, architecture, and data flows, providing reusable implementation patterns for integrating deductive and inductive models.

\begin{figure}[htbp]
    \centering
    \begin{minipage}{0.18\textwidth}
        \centering
        \includegraphics[width=\linewidth]{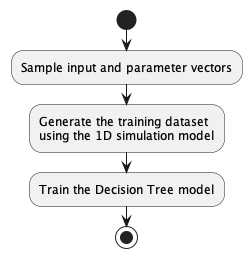}
        \caption*{(a) Creation}
    \end{minipage}
    \hfill
    \begin{minipage}{0.29\textwidth}
        \centering
        \includegraphics[width=\linewidth]{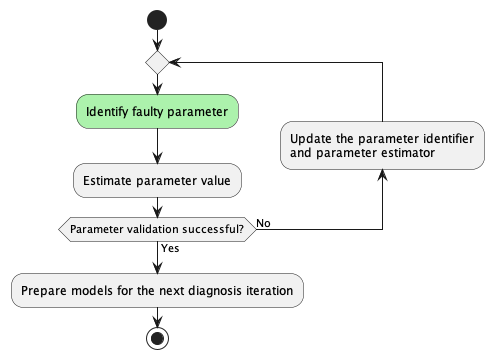}
        \caption*{(b) Usage}
    \end{minipage}
    \caption{Parameter identifier lifecycle}
    \Description{Parameter identifier lifecycle.}
    \label{fig:parameter-identifier-lifecycle}
\end{figure}

\paragraph{Parameter Identification.}
Figure~\ref{fig:parameter-identifier-lifecycle} illustrates the realization of the parameter identification capability. During the creation phase (Fig.~\ref{fig:parameter-identifier-lifecycle}a), a Decision Tree parameter identifier is trained from a dataset generated by the 1D simulation model. During operation (Fig.~\ref{fig:parameter-identifier-lifecycle}b), the trained model identifies the faulty parameter, while a complementary hybrid model estimates its value \cite{wangHybridPhysicsbasedDatadriven2022a}. Following parameter validation, both parameter identifier and estimator models are updated to support subsequent diagnosis (Fig.~\ref{fig:parameter-identifier-lifecycle}b) iterations. Further implementation details are provided in \cite{imhogiemheTHERMALHYDRAULICPROCESSSUPERVISION}.

\section{Results \& Discussion}\label{results-and-discussion}
The proposed Hybridization Characteristics (\textit{HC}) provide a structured way to document both the motivation and realization of hybrid Digital Twins, making explicit design decisions that are often implicit in industrial developments. Applied to the fluidic loop case study, the framework enabled the systematic identification and description of each hybridization capability involved in the diagnostic workflow, including parameter identification, parameter estimation, and parameter validation \cite{datongnon_parameter_identifier_cetim,datongnon_parameter_estimator_cetim,datongnon_parameter_validation,datongnon_hybridization_characteristics_motivation}. This explicit representation improves traceability between engineering needs and their implementations while facilitating knowledge transfer across use cases. Beyond their descriptive role, the realization characteristics reveal recurring implementation patterns that can be generalized across hybrid Digital Twins. For instance, similarly to \cite{zhangPhysicsguidedConvolutionalNeural2020}, both the parameter identifier and the parameter estimator follow the same orchestration pattern: a deductive model is first used to generate a dataset, from which an inductive model is subsequently trained. The recurrence of such workflows suggests that realization characteristics can evolve from documentation artifacts into reusable orchestration templates. This observation opens the perspective of standardizing the orchestration of deductive and inductive models. In the co-simulation domain, the \fmi \cite{fmi_standard} provides a tool-independent standard for exchanging simulation models, while \ssp \cite{ssp_standard} defines how complete systems composed of one or more component models, including FMUs, and their parameterization can be represented and transferred between simulation tools. Building on this separation between model exchange and system-level composition, the realization dimension could similarly provide a standardized means of describing how deductive and inductive models are created, integrated, and interact across both development and runtime phases, thereby facilitating the reuse, interoperability, and engineering of hybrid Digital Twins.

\section*{Conclusion}\label{conclusion}
This paper presented an engineering exemplar of a hybrid Digital Twin developed for a fluidic loop system at CETIM. By extending an existing Digital Twin characterization framework with \textit{Hybridization Characteristics HC}, it provides a structured and traceable description of the motivations and realization of hybridization, explicitly linking engineering needs to implemented solutions.

Beyond documentation, the case study revealed recurring realization patterns that suggest \textit{HC} could evolve toward reusable orchestration patterns for deductive and inductive models. This perspective opens the way to more systematic engineering and future standardization of hybrid Digital Twins.

\newpage
\bibliographystyle{ACM-Reference-Format}
\bibliography{references}

\end{document}